\documentstyle[psfig]{aa}
\def\zabs{$z_{\rm abs}$}
\def\zem{$z_{\rm em}$~}
\def\lya{Lyman-$\alpha$}

\def\ovi{O~{\sc vi}}

\def\nv{N~{\sc v}}
\def\niii{N~{\sc iii}$\lambda$989~ }
\def\ovia{O~{\sc vi}$\lambda$1031~ }
\def\ovib{O~{\sc vi}$\lambda$1037~ }

\def\ciii{C~{\sc iii}$\lambda$977~}
\def\cii{C~{\sc ii}}
\def\ciis{C~{\sc ii*}}
\def\mgii{Mg~{\sc ii}}
\def\civ{C~{\sc iv}}
\def\civa{C~{\sc iv}$\lambda$1548~ }
\def\civb{C~{\sc iv}$\lambda$1550~ }
\def\nva{N~{\sc v}$\lambda$1238~ }
\def\nvb{N~{\sc v}$\lambda$1242~ }

\def\alii{Al~{\sc ii}}
\def\aliii{Al~{\sc iii}}

\def\siiis{Si~{\sc ii*}}
\def\siii{Si~{\sc ii}}
\def\siiii{Si~{\sc iii}$\lambda$1206}
\def\siiv{Si~{\sc iv}}
\def\siiva{Si~{\sc iv}$\lambda$1393~ }
\def\siivb{Si~{\sc iv}$\lambda$1402~ }

\def\mgiia{Mg~{\sc ii}$\lambda$2796~ }

\def\kms{km~s$^{-1}$}

\begin{document}
\thesaurus{11.17.1;11.17.4 APM~08279+5255}
\title{Physical conditions in 
broad and associated narrow absorption-line systems toward
APM 08279+5255
\thanks{Based on
observations collected at the W.M. Keck Observatory, which is operated as a
scientific partnership among the California Institute of Technology, the
University of California and the National Aeronautics and Space
Administration. The Observatory was made possible by the generous financial
support of the W.M. Keck Foundation.}
}
\author{R. Srianand\inst{1}, Patrick Petitjean\inst{2,3}}
\institute{$^1$IUCAA, Post Bag 4, Ganesh Khind, Pune 411 007, India \\
$^2$Institut d'Astrophysique de Paris -- CNRS, 98bis Boulevard 
Arago, F-75014 Paris, France\\
$^3$UA CNRS 173 -- DAEC, Observatoire de Paris-Meudon, F-92195 Meudon
Cedex, France 
}
\date{ }
\offprints{R. Srianand}
\titlerunning {associated absorption toward APM~08279+5255}
\maketitle
\markboth{}{}
\begin{abstract}

Results of a careful analysis of the absorption systems with 
\zabs$\simeq$\zem seen toward the bright, $z_{\rm em}$~$\sim$3.91,  
gravitationally lensed quasar APM~08279+5255 are presented.  
\par\noindent
Two of the narrow-line systems, at \zabs~=~3.8931 and \zabs~=~3.9135, 
show absorptions from singly ionized species with weak or no \nv~ and
\ovi~ absorptions at the same redshift. 
Absorption due to fine structure transitions of \cii~ and \siii~ 
(excitation energies corresponding to, respectively, 156$\mu$m and 34$\mu$m) 
are detected at \zabs~=~3.8931. Excitation by IR radiation is favored as
the column density ratios are consistent with the shape of APM~08279+5255
IR spectrum. The low-ionization state of the system favors a picture
where the cloud is closer to the IR source than to the UV source,
supporting the idea that the extension of the IR source is larger than
$\sim$200~pc. 
The absence of fine structure lines at \zabs~=~3.9135 suggests that the
gas responsible for this system is farther away from the 
IR source.
Abundances are $\sim$0.01 and 1~$Z_{\odot}$ at \zabs~=~3.913 and 3.8931 
and aluminum could be over-abundant with respect to silicon and carbon 
by at least a factor of two and five. All this suggests that whereas 
the \zabs =  3.8931 system is probably located within 200~pc from the QSO 
and ejected at a velocity larger than 1000~km~s$^{-1}$, the \zabs = 3.9135 
system is farther away and part of the host-galaxy. 
\par\noindent
Several narrow-line systems have strong absorption lines due to \civ, \ovi~
and \nv~ and very low neutral hydrogen optical depths. 
This probably implies metallicities $Z\geq ~Z_\odot$ although firm conclusion 
cannot be drawn as the exact value depends 
strongly on the shape of the ionizing spectrum.  
\par\noindent
The \civ~ broad absorption has a complex structure with
mini-BALs (width $\le 1000$~\kms) and narrow components superposed on a 
continuous absorption of smaller optical depth. 
The continuous absorption is much stronger in \ovi~ indicating that the 
corresponding gas-component is of higher ionization than the
other components in the flow and that absorption structures in the BAL-flow 
are mainly due to density inhomogeneities.
There is a tendency for mini-BALs to have different covering 
factors for different species. It is shown that a few of the 
absorbing clouds do not cover all the three QSO images, especially we
conclude that the \zabs~=~3.712 system covers only image C.\par\noindent
Finally we identify 
{\sl narrow} components within the BAL-flow with velocity separations
within 5~km~s$^{-1}$ of the \ovi, \nv~ and \siiv~ doublet splittings
suggesting that
line driven radiative acceleration is an important process to explain the
out-flow. 
 
\end{abstract}

\section{Introduction}
The gravitationally lensed
high-redshift Broad Absorption Line (BAL) QSO APM~08279+5255
has been given tremendous interest since its discovery by
Irwin et al. (1998) as it is one of the most luminous objects in the
universe even after correction for gravitational amplification.
Based on the position of the emission lines,
Irwin et al. (1998) derived a redshift $z_{\rm em}$~=~3.87. 
A probably better estimate of the systematic redshift
comes from the detection of CO(4--3) emission 
at $z_{\rm em}$~=~3.9114$\pm$0.0003 by Downes et al. (1999). 

Imaging of the field reveals two main components
(Irwin et al. 1998, Ledoux et al. 1999) separated 
by 0.378$\pm$0.001~arcsec as measured on HST/NICMOS data
(Ibata et al. 1999)  
and of relative brightness $f_{\rm B}/f_{\rm A}$~=~0.773$\pm$0.007. 
The HST images reveal the presence of a third
object C with $f_{\rm C}/f_{\rm A}$~=~0.175$\pm$0.008,
located in between A and B and almost aligned with them. The PSF fits on
the three objects are consistent with the three components being
point-sources and the colors are similar within the uncertainties
suggesting that C is a third image of the quasar (Ibata et al. 1999, 
Egami et al.  1999).
A high-resolution high signal-to-noise ratio spectrum of APM
08279+5255, covering the wavelength  range 4400--10000\AA~  was
obtained using the Keck telescope and made available to the Astronomy
community for analysis (Ellison et al. 1999a,b).
This spectrum, though complicated by the combination of light traveling 
along three different sight lines, is a unique laboratory for studying
the intervening and associated absorption systems. 

It is well known that the origin of associated systems (systems with 
$z_{\rm abs}$~$\sim$~$z_{\rm em}$)
cannot be inferred  directly from their position in the spectrum.
Indeed, absorption can arise from (i) gas ejected by the central engine
at velocity as high as 60000~km~s$^{-1}$ and nonetheless physically
located very close to the source of ionizing photons (e.g. 
the \zabs = 2.24 "mini-BAL" towards Q~2343+125 at \zem~=~2.515; 
Hamann, Barlow \& Junkkarinen 1997a)  or from (ii) gas associated with 
the host-galaxy or with members of a galaxy cluster surrounding the quasar. 
The distinction can be made in terms of physical properties. The systems
belonging to the first class are characterized by high metal enrichment, 
high-ionization parameters, broader line profiles, partial coverage and  
time variability (Barlow et al. 1992, Petitjean et
al. 1994, Savaglio et al. 1994, Hamann 1997, Hamann et al. 1997b,
Barlow \& Sargent 1997, Ganguly et al. 1999, Papovich et al. 1999). 
The second class of
absorber is characterized by classical properties of intervening
systems such as low metallicities (typically 0.01 to 0.1 of solar) and
undisturbed kinematics.

In this study we investigate the nature and physical properties of 
\zabs$\simeq$ \zem systems toward APM~08279+5255. 
In Section 2, we describe the
data and the grids of photoionization models we
use to infer ionizing conditions in the absorbing gas. We
analyse a probably intervening metal line system very close to the emission
redshift in Section 3. The nature of narrow-line systems with
low-ionization conditions are investigated in Section 4. Section 5 describes
the high-ionization narrow-line systems. In Section 6, we
analyse the nature of the BAL outflow 
and in Section 7 we suggest the presence of 
"line-locking" among narrow components in the BAL flow. 
A summary is given in Section 8.

\begin{figure*}
\centerline{\vbox{
\psfig{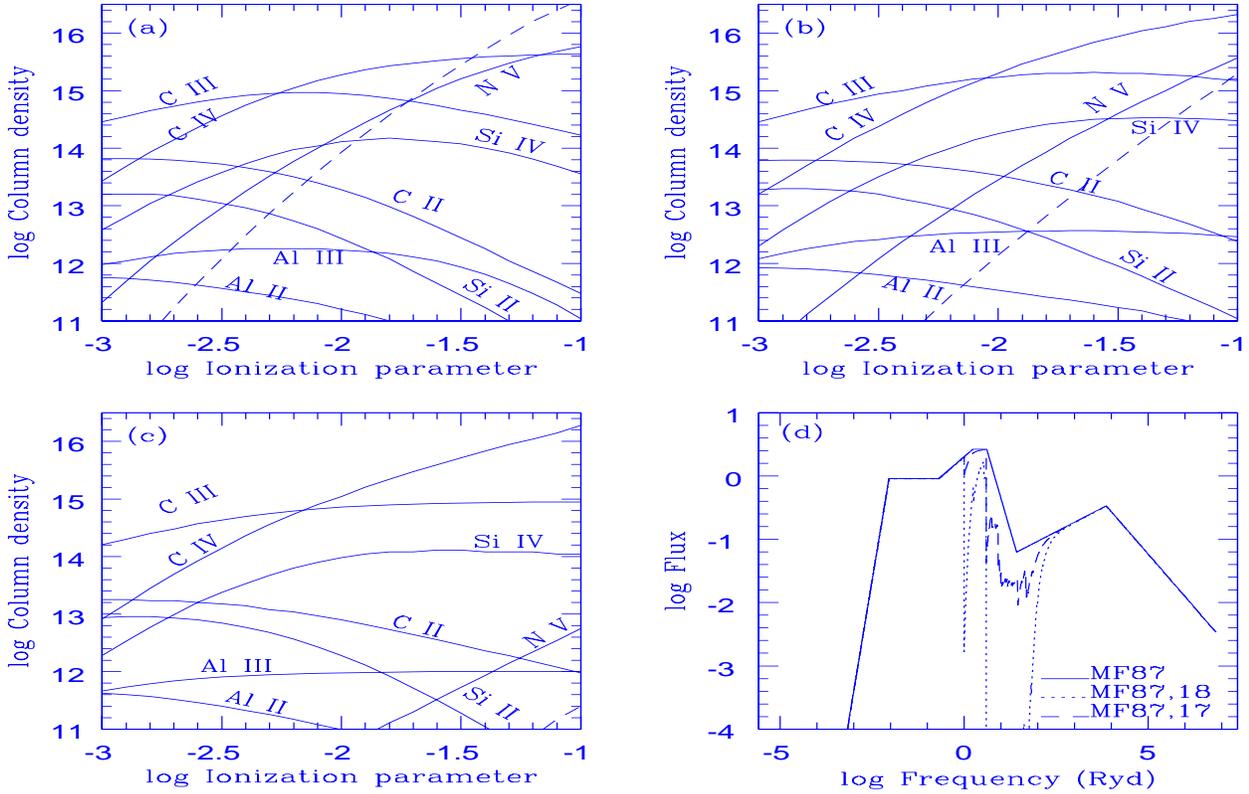}
}}
\caption[]{Results of photoionization models with $Z=Z_\odot$,
log~$N$(H~{\sc i}) = 16.0, constant density and plane parallel
geometry. The dashed line shows the column density of O~{\sc vi}. 
For models in panel(a), 
the Mathews \& Ferland (1987) spectrum has been used;
the same ionizing spectrum has been attenuated by a slab of 
gas with perpendicular neutral hydrogen column density
log~$N$(H~{\sc i}) = 17.0 and log~$N$(H~{\sc i}) = 18.0
and ionization parameter $U$~=~0.1
in panels (b) and (c) respectively. Panel (d) gives the ionizing
spectra in each case: solid line is the Mathews \& Ferland
spectrum; dashed and dotted lines are the attenuated spectra for 
log~$N$(H~{\sc i}) = 17.0 and 18.0 respectively.
}
\label{lis}
\end{figure*}

\section{Data and photoionization models}

A spectrum of  the $z_{\rm em}$~=~3.91
quasar APM 08279+5255  was obtained with the HIRES echelle
spectrograph  at the 10m Keck-I telescope (Ellison et al.  1999a,b).
This data  was made public  together with  a low-resolution spectrum
of  the quasar and a  high-resolution spectrum of  a standard star.  We
have  corrected the high-resolution spectrum of  APM~08279+5255 for
small discontinuities   in  the  continuum,   which  are   probably
due   to  the inappropriate merging  of different orders. These
discontinuities have been recognized by comparing the high  and
low-resolution spectra.  The latter has also  been used  for  normalization
of the  high-resolution  data.   Atmospheric absorption features were
identified from the standard star spectrum.  We  have  measured the
final  spectral  resolution  by fitting  the  narrow atmospheric
absorption   lines  which  are   free  of  blending.   We  find
$FWHM$~$\sim$~8~km~s$^{-1}$ ($b$~$\sim$~4.8~km~s$^{-1}$) at 6900~\AA,
$R$~=~37500, and use this value throughout the paper.

Grids of photoionization models using the code Cloudy (Ferland 1996)
have been constructed. The cloud is a plane parallel slab
of uniform density, solar chemical composition and neutral hydrogen
column densities $N$(H~{\sc i})~=~$10^{16}$~cm$^{-2}$ (see below)
photo-ionized by the QSO radiation. The spectral energy
distribution of APM08279+5255 is not known in the UV/X-ray energy
range. We use the standard AGN spectrum provided by Mathews
\& Ferland (1987) and also consider the effect of screening the
ionizing spectrum by optically thick clouds to take into account
the fact that BALs are X-ray quiet (Green \& Mathur 1996). The resulting 
column densities of various species, for different ionizing spectra, along a
line-of-sight perpendicular to the slab are given in Fig.~\ref{lis}.

\section{The intervening system at $z_{\rm abs}$~=~3.8576}
\begin{figure}
\centerline{\vbox{
\psfig{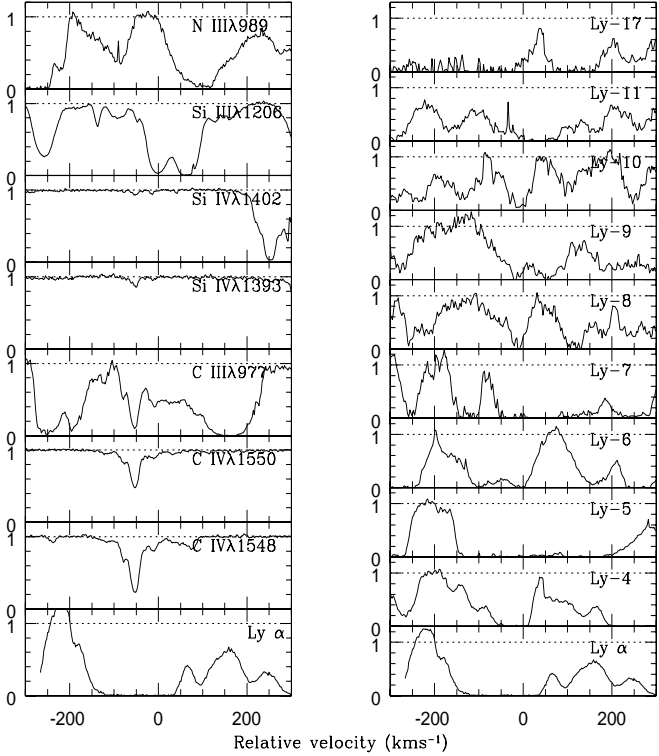}
}}
\caption[]{Profiles of some of the absorption lines at \zabs $\sim$ 3.8576
on a velocity scale centered at this redshift. }
\label{fig1a}
\end{figure}

This system is revealed by strong \civ~ and hydrogen 
Lyman series lines. It is at higher redshift than the main component 
of the BAL out-flowing gas. 
Moderately saturated \ciii~ and \civ~ absorptions are seen
whereas \siiii~ and \siiv, although present, are weak (see
Fig.~\ref{fig1a}). Both \civ~ and \ciii~
profiles suggest the presence of 5 distinct components.
Only the highest velocity component shows absorption due to 
higher hydrogen 
Lyman series transitions. The expected positions of \ovi~ is in 
the wavelength range of the very strong \ovi~ BAL.
 
Column densities estimated from Voigt profile fitting are given in  
Table~\ref{tab385} for different species. Note that the
\civ~ doublet ratio is consistent with complete coverage of
the background source (see below). The neutral hydrogen column density 
at $z_{\rm abs}$~=~3.8574 is well defined by 
higher Lyman series absorption lines.
If we assume the QSO is the ionizing source we derive log~$U=-2.5$
from the C~{\sc iii} and \civ~ column densities and the absence of 
singly ionized species.
Models with solar metallicities produce C~{\sc iii} and \civ~ column densities 
two orders of magnitude larger than what is observed (see Fig.~\ref{lis}). 
This clearly suggests that metallicities are of the order of 
[C/H]~$\simeq$~0.01~[C/H]$_\odot$, similar to what is usually derived for 
intervening systems. It is likely that this system does not belong to the 
QSO environment.

\begin{table}
\caption{Parameters for the system at \zabs=3.8568}
\begin{tabular}{lccc}
\multicolumn{1}{c}{Species}&\multicolumn{1}{c}{$z$}&
\multicolumn{1}{c}{$N$(cm$^{-2}$)}&\multicolumn{1}{c}{$b$(\kms)}\\
\\
H~{\sc i}  &3.8574&1.32$\pm0.13\times10^{16}$ &17.41$\pm$1.13\\
C~{\sc iii}&3.8557&3.13$\pm0.38\times10^{12}$ &20.85$\pm$1.81\\
           &3.8564&7.95$\pm0.50\times10^{12}$ &12.43$\pm$0.36\\
           &3.8568&2.08$\pm0.12\times10^{13}$ &10.67$\pm$0.13\\
           &3.8571&2.41$\pm0.34\times10^{12}$ &~7.25$\pm$0.70\\
           &3.8574&1.16$\pm0.07\times10^{13}$ &12.57$\pm$0.72\\
C~{\sc iv} &3.8557&3.34$\pm0.24\times10^{12}$ &20.85$\pm$1.81\\
           &3.8564&1.09$\pm0.03\times10^{13}$ &12.43$\pm$0.36\\
           &3.8568&3.76$\pm0.43\times10^{13}$ &10.67$\pm$0.13\\
           &3.8571&2.67$\pm0.28\times10^{12}$ &~7.25$\pm$0.70\\
           &3.8574&6.39$\pm0.30\times10^{12}$ &12.57$\pm$0.72\\
Si~{\sc iv}&3.8567&1.27$\pm0.07\times10^{12}$ &~9.73$\pm$0.67\\
           &3.8574&6.69$\pm0.05\times10^{11}$ &~5.44$\pm$0.63\\
\label{tab385}
\end{tabular}
\end{table}
\section{Intrinsic systems with narrow and low-ionization lines}
\subsection{The \zabs = 3.8931 complex}
\begin{figure}
\centerline{\vbox{
\psfig{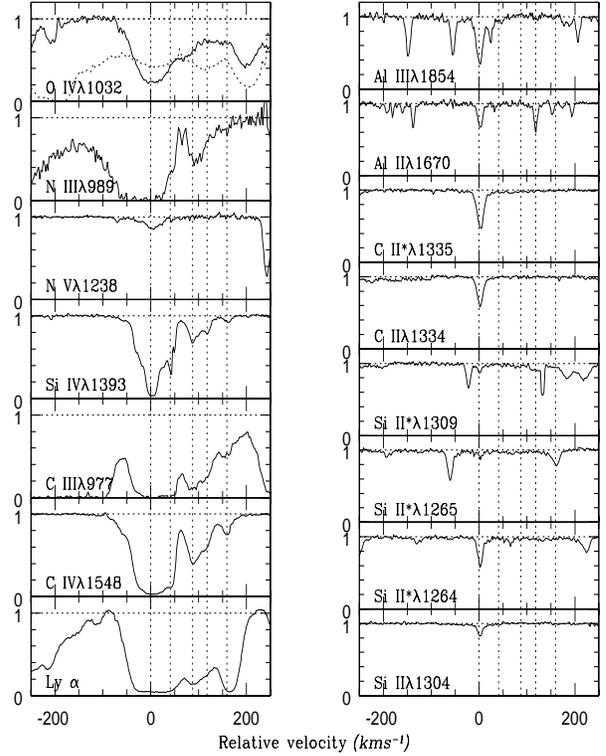}
}}
\caption[]{Profiles of some of the absorption lines at \zabs $\sim$ 3.8931
on a velocity scale centered at this redshift.
The vertical dashed lines mark the position of the components seen in 
the \siiv~ profile. Note that only the component at $v\simeq0$ \kms shows
absorption lines due to singly ionized species. }
\label{fig1}
\end{figure}
Absorption profiles of some of the important transitions from this
complex are given in Fig.~\ref{fig1}. 
In what follows, we concentrate on the
component at $v$~$\sim$~0 \kms, which shows absorption due to low
ionization lines: \alii, \aliii, \siii, \siiis, \cii~ and \ciis.
In addition, there are strong absorption lines at
the expected positions of \niii and C~{\sc iii}$\lambda$977 with
profiles similar to that of
\civ~ and Si~{\sc iv}. It is however
very difficult to rule out contamination by intervening
\lya~ systems. There is a feature at the expected position of 
O~{\sc vi}$\lambda$1031. However, 
as \ovib is heavily blended, it is difficult to establish the presence of
\ovi~ in this system.  \nv~ absorption is present and weak 
(see Fig.~\ref{fig1}). 

\subsubsection{Partial covering factor}
\begin{figure}
\centerline{\vbox{
\psfig{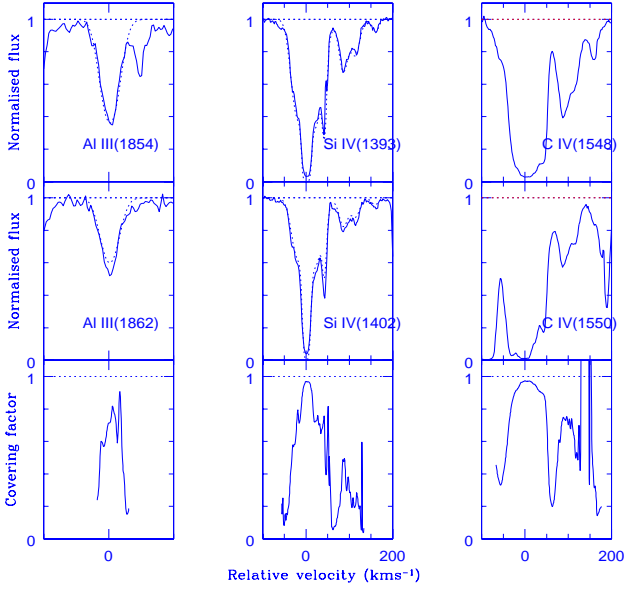}
}}
\caption[]{Covering factor estimated from the Al~{\sc iii}, Si~{\sc
iv} and C~{\sc iv} doublets at \zabs = 3.8931. The top two
rows in each column give the observed profiles. In the
case of \siiv~ and \aliii~ we also plot the best Voigt-profile fit
(dotted lines) for complete coverage to illustrate the inconsistency .
Last row in each column gives the covering factors estimated 
using the residual intensities in each velocity bin.}
\label{fig2}
\end{figure}

\vskip -0.3cm
The flat bottom and the presence of other Lyman series lines show that
the \lya~ line is saturated. However, it is apparent on Fig.~\ref{fig1} that
there is some residual flux in the core of the \lya~ line. 
Similar residuals are seen at the bottom of Si{\sc iv}$\lambda$1393 and 
C{\sc iv}$\lambda$1548. Residual flux is 
expected in case one of the three lines of sight is not intercepted by the 
absorbing cloud. However the relative brightnesses of the components are 
$f_{\rm B}/f_{\rm A}$~$\sim$~0.773 and 
$f_{\rm C}/f_{\rm A}$~$\sim$~0.175 (Ibata et al. 1999). 
Therefore, under the above assumption, at least 10\% residual flux is 
expected, much larger than the 3\% observed. This clearly suggests 
instead that the absorbing gas is located outside the BLR and that it covers 
only 97\% of the background source (BLR+continuum+scattered light). 
We use the \civ, \siiv~ and \aliii~ doublets to estimate the covering factor
of these species applying the method described in 
Srianand \& Sankaranarayanan, 1999. The results are shown in 
Fig.~\ref{fig2}. The covering factor of the \civ~ and \siiv~
component at $v\sim$0~\kms~ is
about 0.95 whereas the corresponding value for \aliii~ is 0.80. 

For simplicity and because the residual is small, we assume
complete coverage while estimating the column density of H~{\sc i}. We
find log~$N$(H~{\sc i})~=~15.9, suggesting that the 
partial Lyman limit detected in the rest frame of the QSO (Irwin et al. 1998) 
is most certainly not due to this system.

\begin{figure}
\centerline{\vbox{
\psfig{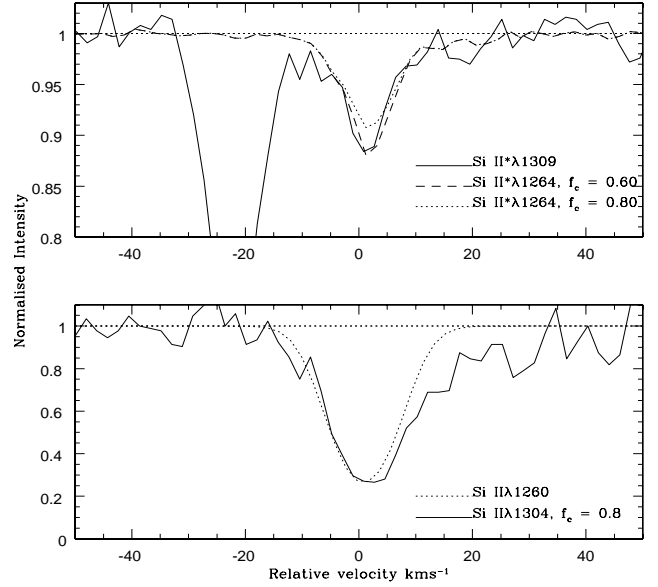}
}}
\caption[]{Covering factor estimated from the \siii~ and \siiis~ lines at
\zabs = 3.8931. {\sl Upper panel}: Absorption profile of Si~{\sc
ii*}$\lambda1309$ (solid line) and profile of Si~{\sc ii*}$\lambda1264$
suitably scaled to be compared to Si~{\sc ii*}$\lambda1264$
for two different values of the covering factor ($f_{\rm c}$~=~0.8 and
0.6 for the dashed and
dotted lines). {\sl Lower panel}: best fit to Si~{\sc ii}$\lambda$1260
(dotted line) after removal of the \civ~ contamination (see text)
over plotted onto the Si~{\sc ii}$\lambda1304$ profile appropriately scaled for
a covering factor of 0.80.}
\label{fig2a}
\end{figure}
The covering factor of the singly ionized phase can be estimated using 
the numerous \siii~ lines. In particular, the Si~{\sc ii}$\lambda$1260
and Si~{\sc ii}$\lambda$1304 absorption lines are redshifted 
beyond the QSO \lya~ emission line. The Si~{\sc ii}$\lambda$1260 line is 
blended with \civa at \zabs = 2.9833 and we use \civb to remove the
contamination. In the lower panel of Fig.~{\ref{fig2a}}, 
the best Si~{\sc ii}$\lambda$1260 fit, after removal of
the \civ~ contribution, is plotted together with the Si~{\sc
ii}$\lambda$1304 profile appropriately scaled to Si~{\sc ii}$\lambda$1260
for a covering factor of $f_{\rm c}=0.8$. This covering factor is consistent 
with the value derived
from the \aliii~ doublet. In the top panel of Fig.~{\ref{fig2a}} 
the Si~{\sc ii*}$\lambda$1309 profile is plotted together with the scaled 
profile of Si~{\sc ii*}$\lambda$1264 for two values of the covering factor. 
The best value is $f\sim0.54$, however as the
Si~{\sc ii*}$\lambda1309$ is weak, a value of 0.8 is well within the
1$\sigma$ error. Thus, the data is consistent with  both \siii~ and
\siiis~ absorptions having similar covering factors $f_{\rm c}\sim0.8$.

The column densities given in the fourth column of Table~\ref{tab389} are 
obtained by integrating the optical depth over the velocity range 
covered by the absorptions after taking into account 
the covering factors derived above. Note that the \civ~ column density 
is unreliable due to blending with other lines
as well as contamination of \civb by \civa at \zabs = 3.901. 
Results from single component Voigt-profile
fits  are given in  the second column of the Table~\ref{tab389}.

\begin{table}
\caption {Parameters for the associated system at \zabs=3.8931}
\begin{tabular}{lccc}
\multicolumn{1}{c}{Ion}&
\multicolumn{1}{c}{$N$(cm$^{-2}$)}&\multicolumn{1}{c}{$b$(\kms)}&
\multicolumn{1}{c}{$N$(cm$^{-2}$)}\\
\\
H~{\sc i}  &8.39$\pm0.37\times10^{15}$ &15.91$\pm$0.57&....\\
C~{\sc ii} &1.64$\pm0.03\times10^{13}$ &8.03$\pm$0.20&2.09$\times10^{13}$ \\
C~{\sc ii*}&2.49$\pm0.04\times10^{13}$ &8.03$\pm$0.20&3.38$\times10^{13}$ \\
C~{\sc iv} &3.58$\pm0.74\times10^{14}$ &8.03$\pm$0.20&2.58$\times10^{14}$\\
Si~{\sc ii}& 7.49$\pm0.34\times10^{12}$ &7.09$\pm$0.40&9.12$\times10^{12}$ \\
Si~{\sc ii*}&2.16$\pm0.30\times10^{12}$ &7.09$\pm$0.40&2.69$\times10^{12}$\\
Si~{\sc iv}& 1.20$\pm0.50\times10^{14}$ &7.09$\pm$0.40&7.76$\times10^{13}$\\
N~{\sc v} &  3.28$\pm0.24\times10^{12}$ &7.09$\pm$0.40&.... \\
Al~{\sc ii}&  6.25$\pm0.15\times10^{11}$ &6.47$\pm$0.19&7.76$\times10^{11}$\\
Al~{\sc iii}& 6.23$\pm0.15\times10^{12}$ &9.07$\pm$0.24&1.04$\times10^{13}$\\
\label{tab389}
\end{tabular}
\end{table}
\subsubsection{Excitation of the fine-structure levels}

The presence of several absorption lines from excited fine-structure levels of
Si~{\sc ii} and C~{\sc ii} is a unique opportunity to study the
physical properties of the absorbing gas. Two main excitation processes 
are at play: radiative excitation by an IR radiation field and collisional
excitation mainly by electrons.

Let us first consider radiative excitation. In that case,
\begin{equation}
{N(X^*)\over N(X)}~=~2~\bigg({\bar n_\lambda \over  1+\bar n_\lambda}\bigg)  
\end{equation}
where $X$ is either C~{\sc ii} or Si~{\sc ii} and 
$\bar n_\lambda$ is given by

\begin{equation}
\bar n_\lambda~=~{I(\nu)\lambda^3\over 8\pi h c}
\end{equation}
where $I(\nu)$, in ${\rm erg~cm^{-2}~s^{-1}~Hz^{-1}}$, 
is the  flux of energy density integrated over all directions.
From the available IRAS fluxes and the sub-millimeter observations by 
Lewis et al. (1999), the flux at the excitation energy of Si~{\sc ii} 
(0.036 eV) in the rest frame of the absorber is 0.1 Jy. 
Assuming the IR emitting region to be a point, $z_{\em}$~=~3.910,
$H_{\rm o}$~=~75~ km~s$^{-1}$ ~Mpc$^{-1}$ and $q_{\rm o}$~=~0.5, we obtain,
\begin{equation}
\bar n_\lambda~=~{0.0038 \over r({\rm kpc})^2} \bigg({1\over k}\bigg)
\end{equation} 
where $r$ is the distance of the cloud from the ionizing source
and $k$ is the magnification factor due to gravitational lensing. From Eqs. (1) and (2) we derive,
\begin{equation}
r({\rm kpc})^2~=~\bigg[ 2{N({\rm Si~{\sc ii }})\over N({\rm Si~{\sc ii}}^*)}-1\bigg]\times {0.0038 \over k}
\end{equation}

The IR flux at the excitation energy of \cii~ (0.008~eV) is 0.016 Jy so that
\begin{equation}
\bar n_\lambda({\rm C~{\sc ii}}^*)~=~15.18\times\bar n_\lambda({\rm Si~{\sc ii}}^*)
\end{equation}
Therefore if radiative excitation is at play, we should observe,
\begin{equation}
{N({\rm C~{\sc ii}})\over N({\rm C~{\sc ii}}^*)}~=~\bigg[{N({\rm Si~{\sc ii}})\over N({\rm
Si~{\sc ii}}^*)} +7.09\bigg]\times0.063.
\end{equation}
As $N$(\siii)/$N$(\siiis)~$\sim$~3.5, the radiative excitation predicts
$N$(\cii)/$N$(\ciis)~$\sim$~0.67 when the observed ratio is 0.66. 
This strongly suggests that indeed excitation by
IR radiation is important. For $k~=~1$, we obtain an
upper limit on the distance of the cloud from the IR source, 
$r$~$<$~150~pc. If the magnification factor is as high as 90, as suggested 
by the probable presence of the third image (Ibata et al. 1999, Egami 
et al. 1999), the distance could be as small as 15~pc.

\begin{figure}
\centerline{\vbox{
\psfig{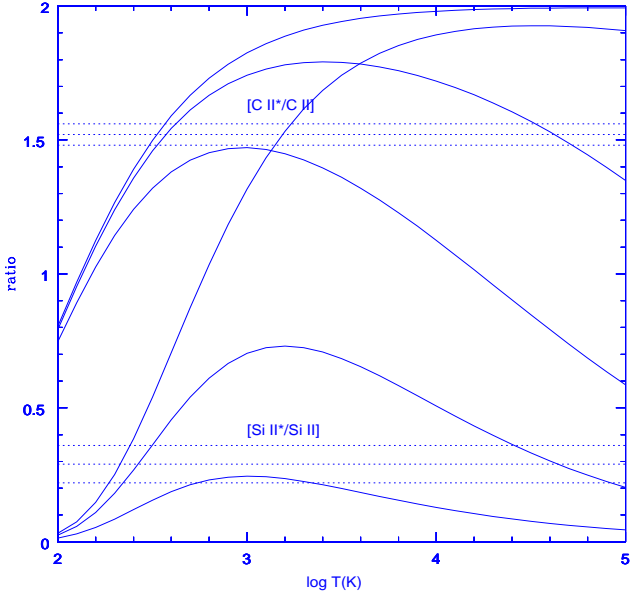}
}}
\caption[]{Column density ratios $N$(X$^*$)/$N$(X) for C~{\sc ii} and
Si~{\sc ii} in the excited and ground states as a function of temperature. 
For each species, the three solid lines correspond to pure collisional 
excitation 
with electronic densities $n_{\rm e}$~=~100, 500 and $10^5$~cm$^{-3}$ 
from the lower to the upper curve. Horizontal dotted lines indicate 
observed values with $\pm$1$\sigma$ errors.
}
\label{fig3}
\end{figure}

Let us consider now the case of pure collisional excitation.
Note that excitation by hydrogen atoms is unimportant in the case of
optically thin clouds. Deexcitation is due to spontaneous emission as
well as particle collisions.
The necessary relationships to estimate the population ratios as a 
function of temperature and density are taken from
Bahcall \& Wolf (1968). For \siii~ the collisional excitation
rate is, ${3.32\times10^{-7} T_4^{-0.5}}~exp({-413.4/T})$~ cm$^3$s$^{-1}$,
the collisional deexcitation rate is, ${1.66\times10^{-7}T_4^{-0.5}}$~
cm$^3$s$^{-1}$ and the spontaneous radiative deexcitation rate is $2.13\times$
$10^{-4}$ s$^{-1}$. For \cii~ the above rates are respectively, 
${6.20\times10^{-8} T_4^{-0.5}}$ $exp({-91.25/T})$~ cm$^3$s$^{-1}$,  
${3.10\times10^{-8}T_4^{-0.5}}$ ~cm$^3$s$^{-1}$ and 
$2.36\times 10^{-6}$ ~s$^{-1}$. Here, $T_4$ is the temperature expressed 
in units of $10^4$~K.

In Fig.~\ref{fig3}, are plotted the ratios $N$(Si~{\sc ii*})/$N$(Si~{\sc ii}) 
and $N$(C~{\sc ii*})/$N$(C~{\sc ii}), of excited to ground state column 
densities as a function of temperature for three different values of the 
electronic density. 
In order to reproduce the observed ratios (indicated by horizontal dashed
lines in Fig.~\ref{fig3}), 
the electron density has to be larger than 100~cm$^{-3}$ and the temperature 
greater than 320~K.  As can be seen from Fig.~\ref{fig3}, there is 
no stringent constraint on the electron density when the electron temperature 
of the gas is low. Note that although it would be surprising to find such a 
small temperature in an optically thin cloud located in the vicinity of the 
quasar, this is possible if the metallicity of the gas is well above solar 
(e.g. Petitjean et al. 1994). In case the gas is warm,
from the width of the absorption lines, we derive an upper limit on the 
temperature of 4.6$\times10^4$~K which gives an upper limit on the density 
of 500 cm$^{-3}$. 

Is there any way to choose between the two processes discussed above?
The small distance inferred in case of pure radiative 
excitation together with the low-ionization state of the gas
may be at odd with the proximity of a very powerful source
of ionizing radiation. From the low spectral resolution spectrum, we estimate the flux at the 
Lyman limit in the rest frame of the absorber to be 
3$\times10^{-16}$~ erg~s$^{-1}$cm$^{-2}$\AA$^{-1}$. 
This corresponds to a luminosity at the Lyman limit of
$L_\nu=4.28\times10^{30}$~ erg~Hz$^{-1}$s$^{-1}$. 
Assuming a flat spectrum and integrating $L(\nu)/h\nu$ over the energy 
range 1 to 20 Ryd, we estimate the number of ionizing photons per 
unit time emitted by the quasar, $Q$~=~2$\times$10$^{57}/k$, where $k$ is
the magnification factor. Thus,
the distance $r$ between the cloud and the quasar is
\begin{equation}
r~=~ {24.3 \over \sqrt{(k n_{100} U_{-2})}}~{\rm kpc}
\end{equation}
where $n_{100}$ is the particle density in units of $10^2$~cm$^{-3}$, 
and $U_{-2}$ is the dimensionless ionization parameter in units of $10^{-2}$.
The latter is the ratio of the density of ionizing photons to the hydrogen
density. The H~{\sc i}, Si~{\sc ii} and C~{\sc ii}
column densities derived in this system are
consistent with log~$U$~$\sim~-$2 (see Table~\ref{tab389} and Fig.~\ref{lis}).
Even for metallicities larger than 10~$Z_{\odot}$, log~$U$~$<~-$1. As 
$n_{100}$~$<$~5 and $k$~$<$~90, we conclude that $r$~$>$~350~pc.  
This lower limit can be
reconciled with the distance inferred from the IR excitation 
if (i) contrary to the UV source, the IR source is not lensed which
does not seem to be the case (see Downes et al. 1999) or (ii) {\sl 
the cloud is closer to the IR source than to the UV source}. Indeed
Downes et al. (1999) have shown that the molecular gas and the dust are
located in a nuclear disk of radius 90 to 270~pc. The absorption system 
could be at 350~pc from the central UV source and still very close to the IR
emitting disk.

Let us now assume that the cloud is at a large distance from
the quasar and that the excitation is purely collisional. 
In that case, the electronic density must be larger than 100~cm$^{-3}$.
Rapid variations of the density on very small scales
could imply variations of the Si~{\sc ii}/Si~{\sc ii}$^*$ ratio from one 
line-of-sight to the other. If we apply a typical ionization correction,
H~{\sc i}/H$^+$~ =~4$\times$10$^{-3}$, to the H~{\sc i} column density 
derived above (see Table~\ref{tab389}), we obtain 
$N$(H)~$\sim$~2$\times$10$^{18}$~cm$^{-2}$ and assuming $n$~$>$~100~cm$^{-3}$,
we infer a dimension of the cloud along the line-of-sight of $\sim$0.006~pc. 
Reverberation studies of the nearby AGN NGC 5548
suggest that the size of the BLR could be in the range 10 
to 20 lt days (e.g. Krolik \& Done 1995, Korista et al. 1995).
It is also known that the size of the BLR scales with  
luminosity as, $L^{0.5}$. From the UV and optical spectra we estimate that 
APM~0827+5255 is at least 500/$k$ times brighter than NGC 5548. 
If we apply the luminosity scaling 
then the radius of the BLR of
APM~0827+5255 is of the order of $\sim$0.3/$\sqrt{k}$~pc. Thus in order 
to cover 97\% of the background source, the transverse dimension of the cloud 
should be $\sim$50/$\sqrt{k}$ times larger than the dimension along the 
line-of-sight. As $k$ could be as large as 100, the two dimensions could be
of the same order of magnitude. The cloud is therefore very small.

\subsubsection{Photoionization model}
In this Section, we discuss  the ionization state of the gas in more detail. 
Fig.~\ref{lis} shows the results of models with solar metallicity 
assuming different
ionizing spectra, either an unattenuated Mathews \& Ferland (1987) spectrum
or the same spectrum attenuated by an ionized hydrogen and helium slab 
with log~$U$~=~--1 and H~{\sc i} column density of 10$^{17-18}$~cm$^{-2}$. 
The resultant spectra are given in panel (d) of Fig.~\ref{lis}.
It is apparent that the absorption is most important at the He~{\sc ii} edge.
This mimics the presence of the 
BAL outflow along the line-of-sight. As an illustration, 
the number of ionizing photons is reduced by a factor 6 due to
the presence of a screen with log~$N$(H~{\sc i})~=~18.

In models with unattenuated ionizing radiation, the 
observed $N$(\aliii)/$N$(\alii),~ $N$(\siiv)/$N$(\siii) and
$N$(\civ)/$N$(\cii) ratios suggest consistently log~$U~=~-2.0$, $-2.2$ and
$-2.3$ and
[Al/H] = 3.2 [Al/H]$_\odot$, [C/H]~=~0.5 [C/H]$_\odot$ and 
[Si/H] = 1.6 [Si/H]$_\odot$. 
Models with 
ionizing radiation attenuated by a screen with 
$N$(H~{\sc i}) = $10^{17}$~cm$^{-2}$ 
give log~$U~=~-2.2$, 
and [Al/H] = 1.7 [Al/H]$_\odot$, 
[C/H]~=~0.5 [C/H]$_\odot$ and [Si/H] = 0.9 [Si/H]$_\odot$. 

The most important effect of attenuation is that the
\nv~ column density decreases with increasing attenuation
due to the paucity of photons of energy larger than the 
N~{\sc iv} ionization potential. In the case of unattenuated ionizing
spectrum and assuming solar abundances, the model predicts 
log~$N$(N~{\sc v})$> 13.2$ which is at least a factor of 3 larger than 
the observed value. Attenuation by a slab with 
neutral column density as large as log~$N$(H~{\sc i})~=~18
is ruled out, however, as in that case log~$N$(N~{\sc v})~=~11 instead of 
12.7 observed. 

In conclusion, if the shape of the ionizing radiation from the
QSO is similar to the Mathews \& Ferland (1987)
spectrum, then the data is consistent with attenuation of the ionizing
flux by a screen with log~$N$(H~{\sc i})~$\sim$~17.
Absolute metallicity is probably very 
close to solar which is typical of associated absorption systems
(e.g. Hamann \& Ferland 1993; Petitjean et al. 1994; 
Petitjean \& Srianand 1999, Hamann \& Ferland 1999 for a recent review). 
However there is an indication that 
Carbon is under-abundant and Aluminum over-abundant compared to Silicon. 

\subsection{The \zabs = 3.9135 complex}
\begin{figure}
\centerline{\vbox{
\psfig{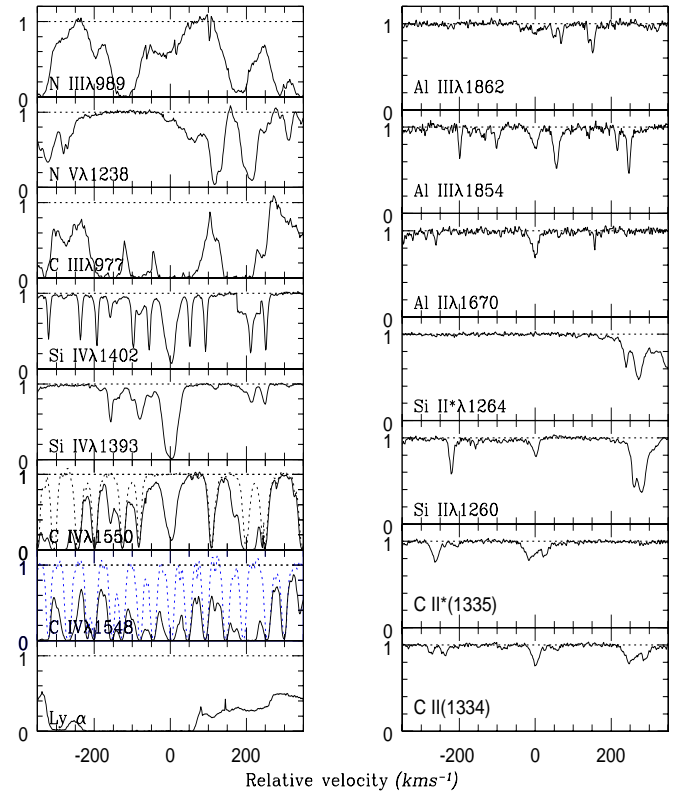}
}}
\caption[]{Absorption profiles at \zabs $\sim$ 3.9135. 
The standard star spectrum is over plotted (dotted line)
to indicate the position of atmospheric absorption features.
}
\label{fig4}
\end{figure}
\begin{table}
\caption{Parameters for the associated system at \zabs=3.9135}
\begin{tabular}{lccc}
\multicolumn{1}{c}{Species}&\multicolumn{1}{c}{z}&
\multicolumn{1}{c}{$N$(cm$^{-2}$)}&\multicolumn{1}{c}{$b$(km/s)}\\
\\
H~{\sc i}  &3.9135&1.04$\pm0.15\times10^{17}$ &18.57$\pm$0.77\\
C~{\sc ii} &3.9135&1.28$\pm0.02\times10^{13}$ &12.51$\pm$0.27\\
C~{\sc ii*}&3.9135&$\le5.30\times10^{12}$ & ....\\
C~{\sc iv} &3.9133&8.13$\pm0.20\times10^{13}$ &11.67$\pm$0.47\\
           &3.9135&4.83$\pm0.17\times10^{13}$ &16.49$\pm$0.49\\
           &3.9138&1.53$\pm0.13\times10^{13}$ &16.71$\pm$3.07\\
Si~{\sc ii}&3.9135&1.35$\pm0.07\times10^{12}$ &11.50$\pm$0.70\\
Si~{\sc ii*}&3.9135&$\le8.40\times10^{11}$ & ....\\
Si~{\sc iv}&3.9133&3.69$\pm0.10\times10^{13}$ & 9.48$\pm$0.26\\
           &3.9135&2.04$\pm0.04\times10^{13}$ &12.70$\pm$0.21\\
           &3.9138&4.53$\pm0.22\times10^{13}$ &10.03$\pm$0.49\\
N~{\sc v}  &3.9135 &$\le10^{12}$               &             \\
Al~{\sc ii}&3.9135 &8.64$\pm0.03\times10^{11}$ &12.21$\pm$0.55\\
Al~{\sc iii}&3.9135&2.15$\pm0.06\times10^{12}$ &12.68$\pm$0.45\\
\label{tab391}
\end{tabular}
\end{table}

The redshift of this system is very close to the CO emission
redshift, $z_{\rm CO}$~=~3.9114 (Downes et al. 1999);
the difference is only $\sim$130~km~s$^{-1}$ whereas the width of the
CO line is 480~km~s$^{-1}$ and the total width of the 
saturated part of the associated Lyman-$\alpha$ line is 300~km~s$^{-1}$.
Three distinct sub-systems are seen in \siiv~ and \civ.
The strongest component in this complex has associated absorption
from \cii, \civ, \siii, \siiv, \alii, \aliii~ and \lya~ (see Fig.~\ref{fig4}). 
Unlike what is observed at \zabs = 3.8931, the \lya~ line goes 
to zero indicating complete coverage. This is confirmed by the \siiv~ and 
\aliii~ doublets. Using the Lyman series, we estimate the neutral hydrogen 
column density, log~$N$(H~{\sc i})~=~17.0$\pm$0.14 (see Table~\ref{tab391}). 
This is consistent with this system being responsible for the
partial Lyman limit absorption seen in the low-resolution spectrum
of APM~08279+5255. The \siiv~ and \civ~ doublets are well fitted by
a model with three components; all the absorption lines due to singly ionized 
species are fitted with a single component. \nv~ and \ovi~ are clearly
absent. There is no absorption from \siiis~ down to a 2$\sigma$ upper limit 
of log~$N$(\siiis)~$<$~11.9.  
The \ciis~ line is redshifted at the same position as \civa at
\zabs = 3.2388. We estimate a 2$\sigma$ upper limit on $N$(\ciis) 
after subtraction of the \civ~ profile, log~$N$(\ciis)~$<$~12.7. The ratio 
of \ciis~ to \cii~ in this system is at least a factor 4 smaller than in 
the \zabs = 3.8931 system discussed previously. This suggests that this cloud 
has to be farther away from the IR source than the \zabs~=~3.8931 cloud. For 
electron temperatures less than 3$\times 10^4$~K the observed limit also
suggests that the electronic density is less than 20~cm$^{-3}$.

The $N$(\alii)/$N$(\aliii) column density ratio in the two systems at 
\zabs =  3.8931 and 3.9135 can be used to compare their ionization states. 
$N$(\alii)/$N$(\aliii)~=~0.40 at \zabs = 3.9135 while the allowed range at 
\zabs = 3.8931 is 0.06--0.10 (after taking into account covering factor 
effects). Note that the $N$(C~{\sc ii})/$N$(C~{\sc iv}) ratio is also larger 
at \zabs = 3.9135. From Fig.~7 of Petitjean et al. (1994) it is clear that when
$N$(H~{\sc i})~$<$~10$^{17}$~cm$^{-2}$, $N$(\alii)/$N$(\aliii)
is approximately independent of the neutral hydrogen column density
(for a given ionizing spectrum) and depends only on the ionization
parameter. Therefore, the ionization parameter of the \zabs = 3.9135 system is
slightly smaller than that of the \zabs = 3.8931 system. 

To discuss this system in more detail, we can use the models of Fig.~1b,
scaling the column densities with abundances.
The $N$(\aliii)/$N$(\alii), $N$(\cii)/$N$(\civ) and
$N$(Si~{\sc iii})/$N$(\siiv) ratios suggest, log~$U=-2.83, -2.6$ and $-2.3$
respectively. The range of ionization parameters reflects the uncertainties. 
If we use the mean ionization parameter, log~$U~=~-2.5$, the model
suggests [C/H] = 0.02~[C/H]$_\odot$,
[Si/H]~=~0.01~[Si/H]$_\odot$ and [Al/H]~=~0.23~[Al/H]$_\odot$. Though
the average abundance in this system is similar to intervening
systems at large redshift, it seems that Aluminum is largely enhanced
compared to Carbon and Silicon. Such peculiar abundance pattern has
already been noted by Ganguly et al. (1999).

All this suggests that whereas the \zabs =  3.8931 system must be
located within 200~pc from the QSO and ejected at a
velocity larger than 1000~km~s$^{-1}$, the \zabs = 3.9135 system must
be farther away and part of the host-galaxy. 

\section{High-ionization narrow-line systems}

In this section we study the high-ionization narrow-line associated systems,
defined as well detached systems with
\lya~ apparent optical depths much smaller than that of \ovi~ and \nv~
(see Fig.~\ref{fig5}). There are two main complexes 
at \zabs$\simeq$3.90 and \zabs$\simeq$3.917. 
We concentrate on well defined strong components.
%
\subsection{ \zabs = 3.8997}
Column densities are estimated
using the apparent optical depth method (Savage \& Sembach 1991).  
The results are given in Table~\ref{tabhigh} together with the velocity 
intervals over which the column density is estimated.
The residual intensities in both components of the \nv~ doublet
and the \ovia line are consistent with complete coverage. 
Unlike \ovi, \nv~ lines are not heavily saturated. As can be seen from  
Fig.~\ref{fig5}, the \civb component is heavily affected by atmospheric 
absorption. We estimate a lower limit on the \civ~ column density using the 
\civa profile and assuming complete coverage. 
There is a strong absorption feature at the expected 
position of \ciii, the profile of which is very similar to that 
of \civ. As we cannot rule out contamination by intervening 
\lya~ absorption, the C~{\sc iii} column density should be considered 
as an upper limit. As can be seen from Fig.~\ref{fig5}, the associated \lya~
absorption is very weak in this component. In addition, this line falls in 
the wavelength range over which \nv~ BAL absorption is present and
coincides with the expected position of the \nvb
absorption from a complex at \zabs = 3.795, which has ubiquitous
absorption due to \civ~ and \siiv~ (see also next Section). 
Therefore the H~{\sc i} column density given in Table~\ref{tabhigh} 
is most certainly an upper limit.
\vskip -0.3cm
\subsection{\zabs = 3.9010}
\begin{figure}
\centerline{\vbox{
\psfig{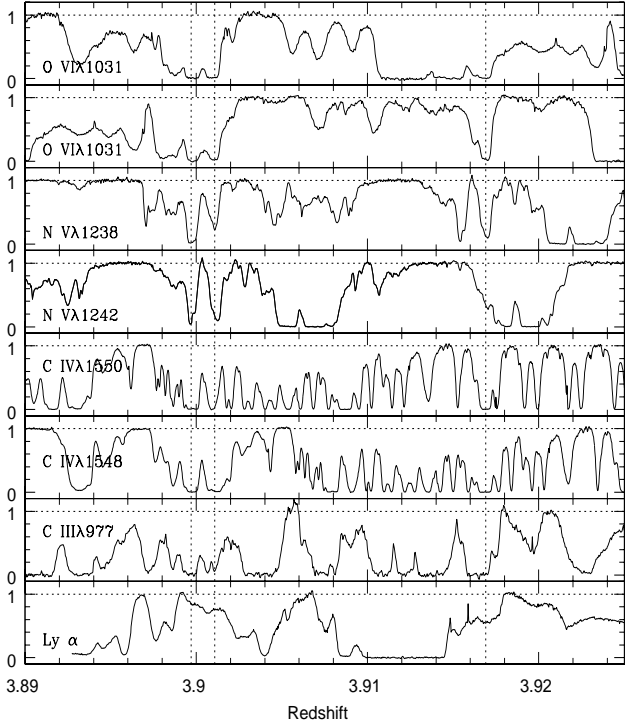}
}}
\caption[]{High-ionization narrow-line systems. The
vertical dashed lines mark the strong components discussed in 
the text. The CO emission redshift is $z_{\rm em}$~=~3.9114.}
\label{fig5}
\end{figure}

%
This component shows strong
\ovi~ absorption with profiles consistent with complete coverage. 
The \lya~ line coincides with the expected position of the \nvb absorption 
from the complex at \zabs = 3.795. Thus, the H~{\sc i} column
density given in Table~\ref{tabhigh} is an upper limit. 
From Fig.~\ref{fig5}, it can be seen that \nvb 
from this system coincides with \nva at \zabs = 3.9170. Thus we estimate 
the \nv~ column density using the \nva line only. As noted before, \civb 
is heavily affected by atmospheric absorption; \civa coincides with \civb 
at \zabs = 3.9135 (see Figs.~\ref{fig4},\ref{fig5}) preventing us to
estimate the \civ~ column density in this system. There is a strong line at 
the expected position of C~{\sc iii}$\lambda$977. We estimate an upper limit 
of the C~{\sc iii} column density assuming complete coverage.
\vskip -0.3cm
\subsection{\zabs = 3.9170}
The saturated \ovi~ lines have no residual flux so 
the absorbing cloud completely covers the background source. The \ovib line
is blended with a strong line most probably due to an intervening
cloud. The lower limit on the \ovi~ column density is estimated using the 
\ovia line only. As discussed before the \nva line is blended with the \nvb 
line at \zabs = 3.9010 and the \nvb line is blended with \mgiia at \zabs
= 1.18. It is therefore difficult to estimate the \nv~ column density
accurately. We obtain an upper limit using the \nva profile. Though the \civ~ 
absorption is quite strong, both lines of the doublet are badly blended with 
atmospheric features.
\begin{table}
\caption{Parameters for the high-ionization narrow line systems}
\begin{tabular}{clrc}
\multicolumn{1}{c}{\zabs}&\multicolumn{1}{c}{Species}&
\multicolumn{1}{c}{$N$(cm$^{-2}$)}&\multicolumn{1}{c}{$v$(\kms)}\\
\\
3.8997&H~{\sc i}&$\le3.7\times10^{12}$& -20,18\\
      &\nv      &2.85$\times10^{14}$& .....\\
      &\civ     &$\ge2.3\times10^{14}$& .....\\
      &C~{\sc iii}&$\le1.0\times10^{14}$& .....\\
      &\ovi     &$\ge9.8\times10^{14}$& .....\\
3.9010&H~{\sc i}&$\le6.3\times10^{12}$& -14,18\\
      &\nv      &6.91$\times10^{13}$& .....\\
      &C~{\sc iii}&$\le3.0\times10^{13}$& .....\\
      &\ovi     &$\ge7.3\times10^{14}$& .....\\
3.9170&H~{\sc i}&2.60$\times10^{13}$& -50,18\\
      &\nv      &$\le1.0\times10^{14}$& .....\\
      &\ovi     &$\ge9.7\times10^{14}$& .....\\
\label{tabhigh}
\end{tabular}
\end{table}

\subsection{Modeling}
The column densities derived in the three systems are given in 
Table~\ref{tabhigh}. The observed column densities of H~{\sc i}, \civ, \nv~ 
and \ovi~ are typical of associated absorption systems (Petitjean et al. 1994, 
Hamann 1997, Petitjean \& Srianand 1999). 

We run photoionization models using
the "Optimize" command available in "Cloudy" to derive the best
values for the ionization parameter and heavy element abundances.
For the Mathews \& Ferland (1987) ionizing spectrum we
obtain log~$U=-0.87$ and $Z=180Z_\odot$ for the \zabs = 3.8997 system and 
log~$U=-0.74$ and $Z$(O,N)~=~14$Z_\odot$ for the \zabs = 3.9010 system. 


However, it is known that BALQSOs are X-ray quiet with optical
to X-ray spectral index $\alpha_{\rm OX}\ge 1.9$ (Green \& Mathur,
1996) while the Mathews \& Ferland spectrum has 
$\alpha_{\rm OX}~=~1.40$.    
If we assume the spectra seen by the absorbing
clouds to be steeper than the  Mathews \& Ferland spectrum then we obtain
$Z$~=~5, 1.4, 0.5, 0.4~$Z_{\odot}$ and $U$~=~0.3, 0.7, 1.2, 1.3 
for $\alpha_{\rm OX}$~=~1.5, 1.7, 1.9 and 2.1 respectively for the
\zabs = 3.8997 system.
The reason for this is that 
the number of photons available for ionization of C~{\sc iv}, \nv~ and
O~{\sc vi} decreases when $\alpha_{\rm OX}$ increases. 
Therefore, it would be of first interest to measure $\alpha_{\rm OX}$
for this quasar.

\begin{figure*}
\centerline{\vbox{
\psfig{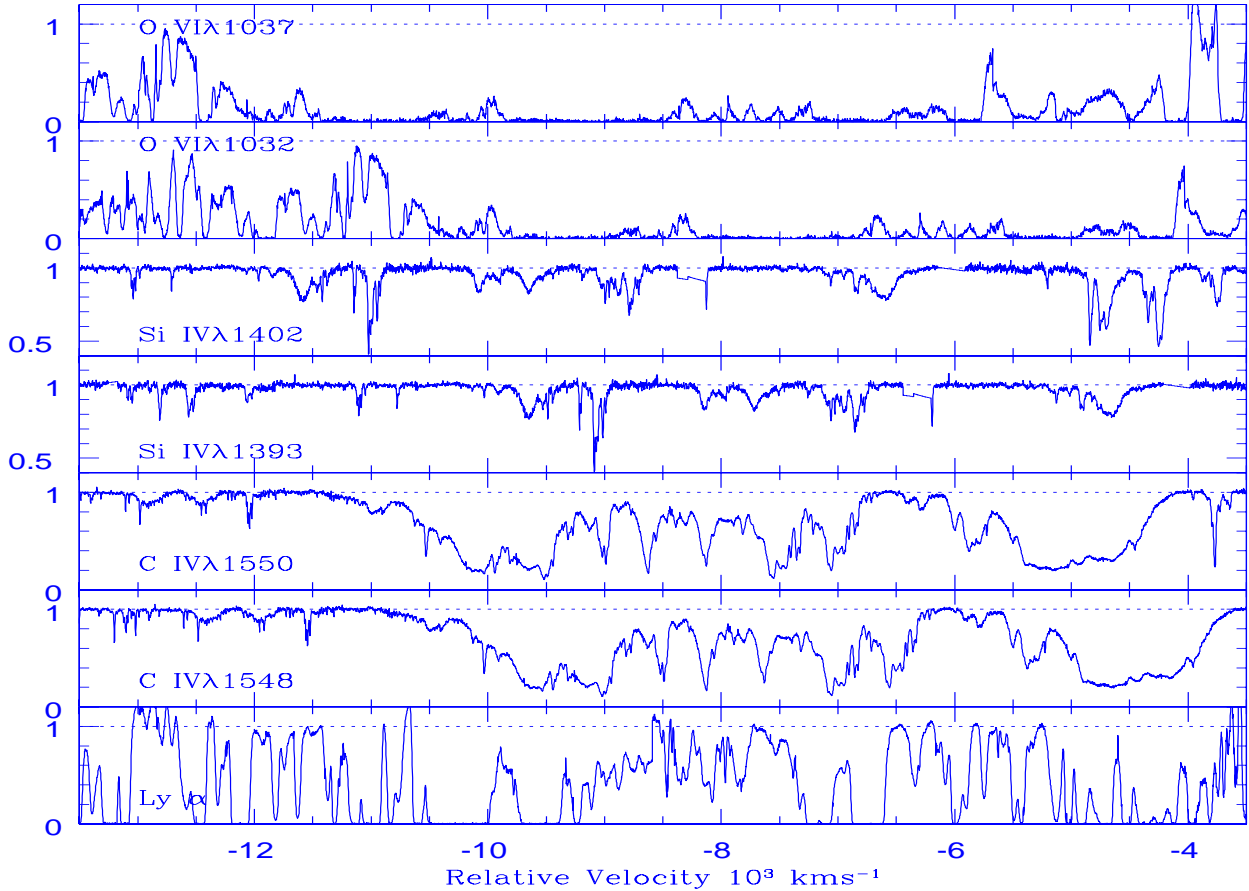}
}}
\caption[]{Broad absorptions due to various species. The origin of
the velocity scale is taken at $z_{\rm em}$~=~3.91.
}
\label{hbal}
\end{figure*}
%


%

\vskip -0.3cm
\section{The BAL flow}

The maximum  observed ejection velocity of the C~{\sc iv} absorption is $\sim
-12000$~km~s$^{-1}$ relative to the systemic redshift $z_{\rm em}$~=~3.911.
The BAL is of high-ionization:
the \ovi~ absorption is much stronger than the C~{\sc iv}
one; \nv~ is also conspicuous; weak \siiv~ absorption
is seen in only a few components (especially those responsible for the 
strong \civ~ troughs at $\sim -9670$ \kms~ and $\sim-$4670 \kms).  
The structure of the absorption profiles due to 
the out-flowing gas is complex as it breaks into
components of different widths, depths and covering factors
(see Fig.~\ref{hbal}). 

Weymann et al. (1991) have shown that
the mean spectrum of BALQSOs shows two distinct and localized
minima separated by about the velocity splitting of \nv~ and \lya~
(i.e.$\sim5900$~\kms) located near velocities $-$4700 and $-$10500 \kms~ 
relative to the emission redshift. Korista et al. (1993) have found that 
22\% of BALQSOs do show double troughs. These features suggest that 
radiative acceleration is important to generate the flows and the observed 
profiles are due to line-locking (Arav 1996 and references therein).

\begin{figure}
\centerline{\vbox{
\psfig{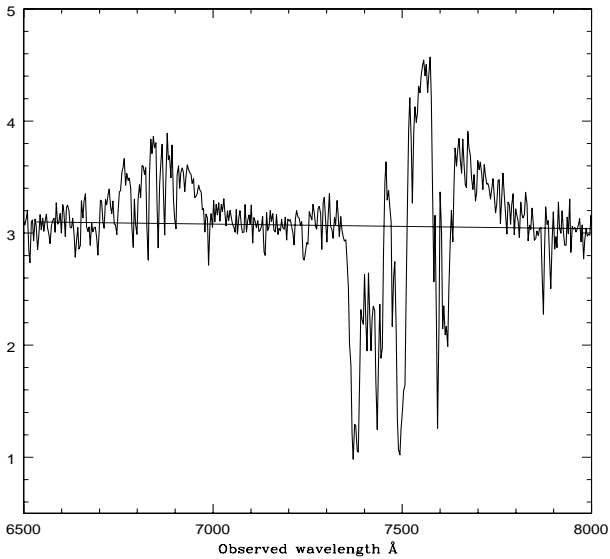}
}}
\caption[]{Part of the low-resolution spectrum showing 
the \siiv~ and \civ~ emission lines. The solid line
indicates the QSO continuum emission.}
\label{lbal}
\end{figure}

In APM~08279+5255, the two strongest troughs in the \civ~ profile are
separated by $\sim 4900$ \kms (see Figs.~\ref{hbal} and \ref{hbal}), 
smaller than the
\lya--\nv~ velocity splitting. In between these two absorption troughs
there are weak \civ~ absorption lines (resolved even in the
low-resolution spectrum). From the shape of the red wing of the \civ~ 
emission line it is also apparent that the flux in the blue wing of the \civ~
emission line at $\sim 7461$\AA~ is very close to the expected 
unabsorbed value (see Fig.~\ref{lbal}). This suggests that the two strong 
troughs are well detached. It is very difficult to produce such a profile in
models of out-flowing gas where the double structures are created by
radiative processes which produce clustering in the velocity
space keeping all absorbing atoms at the same physical location
(see the fits to double troughs in Fig.~11 of 
Arav 1996). We thus believe that the two strong troughs are due to physically 
distinct absorbing components which are similar to mini-BAL systems
seen in a few QSOs (e.g. Petitjean \& Srianand 1999).

In what follows we consider smaller velocity intervals
and investigate the profiles of different absorptions in more detail.
The velocity are taken with respect to $z_{\rm em}$~=~3.911.

\subsection{The broad C~{\sc iv} absorption at $v\simeq-12400$ \kms}
\begin{figure}
\centerline{\vbox{
\psfig{figure=9340.f11,height=9.cm,width=9.cm,angle=0}
}}
\caption[]{Profile of the \civa (bottom) and \civb (middle)
absorption lines from the gas at $v\simeq-12400$ \kms with respect to the
QSO. The normalised spectrum of the standard star is over plotted
to visualize the narrow atmospheric absorption lines.
In the top panel are plotted the \civa (solid) and \civb
profiles (dashed) properly scaled for covering factors
0.10, 0.12 and 0.14. The shaded regions are the velocity
intervals affected by atmospheric absorption. }
\label{balcf1}
\end{figure}

The two broad features at 7295 \AA~ and 7305 \AA~ correspond mostly to
the two lines of a \civ~ doublet at $z_{\rm abs}$~$\sim$~3.712
($v\simeq-12400$ \kms~ in Fig.~\ref{hbal}). The profiles do not match exactly 
however. It is clear from Fig.~\ref{balcf1} that there is an extra absorption 
in the \civb profile at $\sim-160$ \kms~ from the center of the line. 
This absorption is tentatively identified as \civa at
\zabs$\simeq$~3.717, the corresponding broad, albeit weak, \civb line
is blended with the narrower  Fe~{\sc ii}$\lambda$2600 complex from
a \mgii~ system at $z_{\rm abs}$~=~1.81. Most of 
the narrow absorption features seen in the \civa and \civb profiles are due 
to atmospheric absorptions.

We believe, this absorption structure is real because of the good match 
between the profiles of the \civ~ lines in the red wings 
(see Fig.~{\ref{balcf1}}), and because the lines 
are located very near the centre of an echelle order and thus cannot be an 
artifact of imperfect order merging. This system is well detached from the 
rest of the BAL and therefore the continuum can be fitted accurately. 
The \civ~ absorption is most probably saturated as the residual intensities 
in both lines are similar. 
However they are equal to 0.87 which
clearly indicates that the covering factor is surprisingly small, $\sim0.13$. 
The \civ~ profiles, in the velocity range not affected by
the narrow atmospheric absorption lines between $-$100 \kms~ and 100 \kms, 
can be fitted with a covering factor of 0.12 (see top panel in 
Fig.~\ref{balcf1}). The typical error in the covering factor is 0.02 (as the 
rms in the normalised continuum over this wavelength range is 0.02). 
From Fig.~{\ref{lbal}} it is apparent that this absorption, 
does not occur on top the \civ~ emission line. Thus the low coverage
is most likely due to the fact that the absorbing gas does not
cover one of the images of the very small continuum source. 
If the absorbing gas covers only image C,
then we expect a covering factor of 0.09. The consistency 
with the observed value is an additional argument in favor of 
object C being the third image of the gravitationally lensed quasar.
If true, this implies that a strong \civ~ absorption, with very little 
residual intensity, is expected in the spectrum of image C whereas
very weak or no absorptions are expected in the spectra of A and B.   
This can be probed with HST/STIS observations.

The width of the absorption lines is $\sim200$ \kms. This system must be
considered as a "mini-BAL" as  the \civ~ profile is smooth and broader
than a typical intervening system but narrower than a typical BAL
outflow (Barlow et al. 1997). \siiv~ absorption is not detected and \ovi~ 
and \nv~ are redshifted at the same position as other strong lines.
\subsection{Absorptions in the range $-11000<v<-9000$~\kms}
\begin{figure}
\centerline{\vbox{
\psfig{figure=9340.f12,height=10.5cm,width=9.5cm,angle=0}
}}
\caption[] {Absorptions in the velocity range $-10300<v$ $<-9200$ \kms.
The zero of the velocity scale is taken at $z$~=~3.911.
The vertical doted and dashed lines are drawn for illustrative purpose
at velocities discussed in the text.
}
\label{bal1}
\end{figure}

In addition to the broad absorption seen in this wavelength range,
there are a few narrow absorption lines identified as \civ~ doublets.
%
\subsubsection {Narrow lines at $v\simeq-10030$ \kms}
The system at $v\simeq-10030$ \kms~ is defined by \civ, \ovi, \nv~ and \siiv~
narrow absorption lines. They are most probably associated with a 
strong \lya~ line seen at the same velocity (see Fig.~\ref{bal1}). Unlike 
other narrow systems seen over this wavelength range, the profile of the 
\ovi~ absorption is well defined and less affected by blending. 
The equivalent widths of the \civ~ lines measured using a continuum 
defined locally, $w$~=~0.24 and 0.13~\AA, are consistent with a  
covering factor close to unity.

A Voigt-profile fit to the \civa narrow line, with a continuum defined 
locally, gives $b$ = 11.$\pm$0.3~\kms and 
$N$(\civ)~=~$1.5\pm0.1\times10^{13}$~cm$^{-2}$. This gives an upper limit 
on the temperature of the gas, $T\le 9\times10^4$~K.
The \civa, \siiva and \ovib profiles are very similar, suggesting that the 
velocity dispersions are of the same order. Higher order Lyman
series lines are present exactly at the redshift defined by the \civa
line. 
The Ly-6 line is weak and not saturated. Voigt-profile fitting to 
this line gives $N$(H~{\sc i})~=~3.3$\pm$0.2~10$^{15}$~cm$^{-2}$ and 
$b~=~20\pm2$ \kms, which implies $T\le 2.4\times10^4~K$ and suggests 
that ionization is dominated by photoionization. 

%
\subsubsection {System at $v\simeq-9650~$\kms}
This system is defined by strong and broad \civ~ and \siiv~ lines. There
is an absorption line at the expected position of \siiii~ with very similar
velocity profile which leaves little doubt on the identification 
(see Fig.~\ref{bal1}). 
The fit to the \siiv~ doublet gives a covering factor of $\sim$0.25.  
For this value we estimate the column density of \siiv~ to be 
$\sim 10^{14}$~cm$^{-2}$. 
It is interesting to note that the residual intensities of \siiii~ and
\siiv~ are similar. If both absorptions are produced in the same region this
means that the column density of \siiv~ is about a factor of 3 larger than
that of Si~{\sc iii}. 

As the continuum fitting in the \siiv~ region is accurate, the estimate
of the covering factor is reliable. The low-resolution data indicates
that the \siiv~ absorption is not located on top of the \siiv~ emission
line. Thus, a partial covering factor has to be explained in terms of 
one or two of the multiple images not being covered.
However, the covering factor we derive here is larger than what
is expected if the gas covers only image C and smaller than the
expected value if it covers only image B. This most certainly means
that the absorbing gas covers at least two of the images and that the
optical depths along different sight lines are different. Absorptions due
to the \civ~ lines are strong and both have residual intensity
$\sim$0.20 (covering factor $f\sim0.80$). Though this
value is somewhat uncertain due to the non-uniqueness of the continuum 
fitting, it is larger than the value derived for \siiv. 
It has been observed already in some "mini-BALs" that covering 
factors vary from one transition to the other (see e.g. the \zabs = 2.207 
system towards J2233-606, Petitjean \& Srianand 1999).
However, the interpretation of this in the present case is clearly that 
the ratio of the optical depths of \siiv~ and \civ~ along different
sight-lines are different.

The Lyman-$\alpha$ line at
about the same velocity is most certainly from an intervening system at
$z_{\rm abs}$~$\sim$~3.7584 associated with two weak \civ~ and 
\siiv~ narrow components (marked by dotted lines at $-9440$ and $-9530$~\kms~
on Fig.~\ref{bal1}). 
The determination of the H~{\sc i} column density in the broad system
is difficult as the higher Lyman series lines 
are expected to be broad
and shallow if the covering factor is 0.25. However we note that the
residual flux at the expected positions of Ly-6 and Ly-7 are smaller
than 0.25 suggesting that these lines, if at all present, are not 
heavily saturated. We obtain an upper limit on $N$(H~{\sc i}) of
4$\times10^{14}$~cm$^{-2}$ for an assumed covering factor of 0.25. 
For a Mathews \& Ferland (1987) ionizing spectrum, the ionization parameter
required to produce the observed \siiv~ to Si~{\sc iii}~ column density 
ratio is log~$U\simeq-2.00$.
This suggests that the absorbing gas is weakly ionized compared to
the systems we have discussed above. The model needs 
[Si/H]$\ge$20[Si/H]$_\odot$ in order to reproduce the column
densities. Note that in this case the solution depends weakly on 
the assumed value of $\alpha_{\rm OX}$

\subsection{Narrow components in the range $-9600<v<-7000$ \kms}
%

From the \civ~ profiles (see Fig.~\ref{hbal}), it is apparent that this
velocity range is dominated by absorptions from individual clouds 
superposed on a continuous absorption. It is difficult to 
disentangle both contributions. 
The narrow \civ~ systems have small H~{\sc i} optical depth. 
This is characteristic of systems associated with the quasar.

It is interesting 
to note on Fig.~\ref{hbal} that the residual 
flux of \ovib is very small and narrow components are barely visible. 
This indicates that the continuous gas-component is of higher ionization 
than the other components in the flow. 
Absorption due to \siiv~ is not detected over this velocity range. Thus, the 
systems are similar to the high-ionization systems discussed in Section~5.

\subsection {Complex centered at $v\sim-7000$\kms}
\begin{figure}
\centerline{\vbox{
\psfig{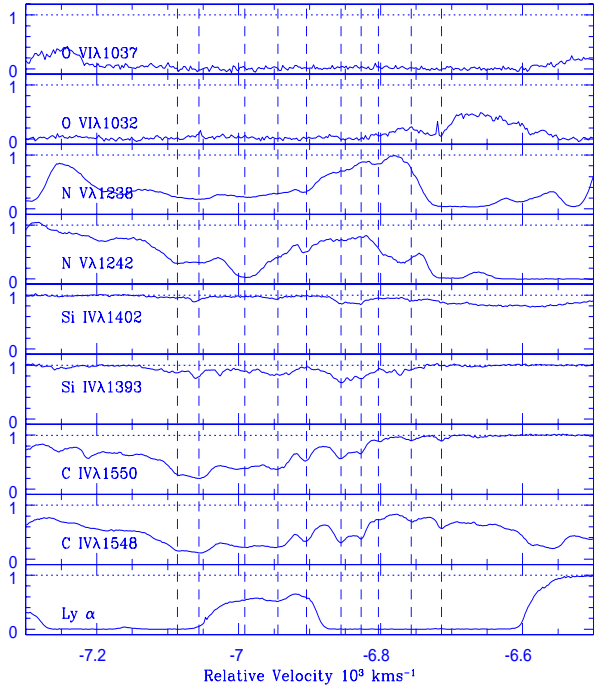}
}}
\caption[]{Absorptions in the velocity range $-7300<v<-6500$ \kms.
The zero of the velocity scale is taken at $z$~=~3.911.
The vertical dashed lines are drawn for illustrative purpose
at velocities discussed in the text.
}
\label{bal3}
\end{figure}
As can be seen in Fig.~\ref{bal3}, there are at least 10 distinct components 
(marked with dashed lines) in the \civ~ profile of this complex. 
The two strong \lya~ lines at $\sim -6800$ and $-7100$ \kms~
are most probably due to intervening systems. The line at 
$v$~=~$-6800$ \kms~ (\zabs = 3.8018) is indeed associated with the 
five weak and narrow C~{\sc iv} and Si~{\sc iv} components with the largest 
velocities ($>-6850$ \kms). 
  
The other \civ~ components have strong associated \nv~ absorption with 
$N$(N~{\sc v}) $>$ $N$(Si~{\sc iv}), typical of associated systems. 
The non similarity of the \nva and \nvb profiles indicates that there is some
blending. As noted before, the expected position of \lya~ from the two
high-ionization systems at \zabs $\sim 3.90$, discussed in Section 5,
coincides with \nvb in this system. 
Moreover, the \nvb line of the system at $v~=~-8120$ \kms coincides with 
the blue wing of the \nva absorption.
The strong absorption in the \nvb profile at $v~=~-6990$ \kms 
is most certainly Lyman-$\alpha$ at $z_{\rm abs}$~=~3.904 with possible
\nv~ but no \ovi~ associated absorption. 
 
From the \nva profile it is clear that the gas producing the \nv~
absorption covers more than 80\% of the background source. Fitting of
the unblended \civ~ and \siiv~ doublets results in covering factors in the 
range 0.8$-$0.9$\pm0.1$. It is clear from Fig.~\ref{bal3} that the apparent  
optical depth of H~{\sc i} in the range $-7050<v<-6900$ \kms~ 
is much smaller than the apparent optical depth of \civ~ and \nv.
This suggests that these systems are very similar to the systems 
discussed in Section~5.

\subsection {The velocity range $-6000<v<-4000$ \kms}
\begin{figure}
\centerline{\vbox{
\psfig{figure=9340.f15,height=10.cm,width=9.5cm,angle=0}
}}
\caption[]{Absorptions in the velocity range $-5700<v<-4200$ \kms.
The zero of the velocity scale is taken at $z$~=~3.911.
The vertical doted lines are drawn for illustrative purpose at 
velocities discussed in the text.
}
\label{bal4}
\end{figure}
The absorption profiles in the velocity range considered here
is mainly due to broad lines (see Fig.~\ref{bal4}). 
%
%
The component at $v~=~-5350$ \kms~
is defined by \civa, \nv~ and \ovi~ absorptions.  
The width of the absorptions is consistent with the definition of "mini-BAL".  
It can be seen on Fig.~\ref{bal4} that the optical depth increases 
dramatically from H~{\sc i} to \ovi~ (note the weak but significant
residual in \ovib). This clearly suggests that
$N$(\ovi)~$>>$~$N$(\civ) and that the ionization of this system is very high. 
%
%

The component at $v~=~-4700$ \kms~
is defined by \civ, \siiv, \nv~ and \ovi~ absorptions.
The two lines of the \civ~ doublet are partially blended. The width of 
the lines is $\sim1500$ \kms. The \siiva profile has a width of $\sim450$ \kms.
The \siivb line is blended with the \civa from \zabs = 3.3776. As the latter 
intervening system  is itself heavily blended, unique profile decomposition 
is not possible and it is difficult to remove the \civa contribution to the 
broad \siivb profile. The important observation is that the residual 
intensities of \ovia, \ovib and \nva are larger than that derived for the 
component at $v~=~-5300$ \kms~ while it is the opposite for
the \civ~ residual intensity. This probably indicates that this system is of
lower ionization compared to the "mini-BAL" at \zabs~=~$-$5300 \kms.
\vskip -0.3cm
\section{Line-locking}
\begin{figure}
\centerline{\vbox{
\psfig{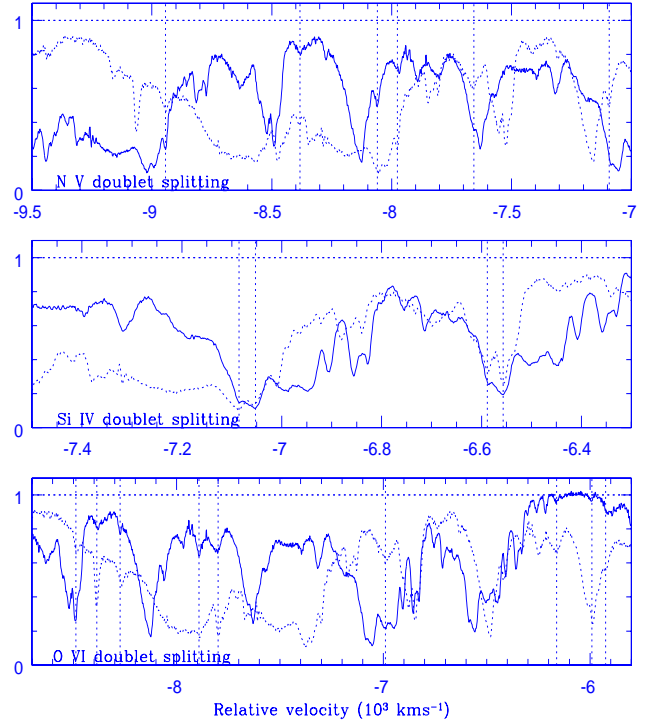}
}}
\caption[]{Velocity separations of narrow absorption lines in the BAL
outflow. The solid line is the observed \civ~ profile.
The dotted line is the \civ~ profile shifted by different doublet splitting: 
\nv, \siiv~ and \ovi~ for the upper, middle and lower panel respectively.
The vertical dashed lines indicate the places where the centroids of lines 
match within 5 \kms.}
\label{bal5}
\end{figure}

Several mechanisms have been proposed to explain the acceleration of 
BAL winds (see a review by de Kool 1997). One of the signatures 
of acceleration dominated by line radiation pressure is the presence
of line-locking (e.g. Q~1303+308 in Foltz et al. 1987; NGC~5548 in
Srianand 2000). 
We searched the spectrum of APM~08279+5255 for such signatures in the \civ~
wavelength range. We note that some of the velocity separations between
the "mini-BALs" are close to the \siiv~ and \ovi~ doublet splittings.
However as the individual lines are broad, the probability of such
occurrence by chance is high. More striking is that most of the narrow 
\civ~ systems have a "companion" system with a velocity difference 
corresponding to the velocity splitting of the \siiv, \nv~ or \ovi~ doublets 
with line centroids coinciding within 5~\kms. 
The maximum number of matchings occurs at the velocity separation of the \ovi~ 
doublet as expected for a high-ionization system. As an illustration we 
show in Fig.~\ref{bal5} the components with velocity separations equal to 
the \ovi, \nv~ and \siiv~ doublet separations. We believe that
the data are indicative of the presence of "line-locking". 
\vskip -0.5cm
\section {Summary}
%
\subsection {Low-ionization systems}

Two of the narrow absorption systems (at \zabs~=~3.8931 and \zabs~=~3.9135), 
show absorptions from singly ionized species (C~{\sc ii}, Si~{\sc ii}, 
Al~{\sc ii}) whereas absorptions from \nv~ and \ovi~ are weak or absent. 
Absorptions from excited fine structure levels of \cii~ (excitation energy
corresponding to 156$\mu$m) and \siii~ (excitation energy corresponding to 
34$\mu$m) are detected at \zabs~=~3.8931. The relative column densities are 
consistent with the shape of the IR spectrum of APM~08279+5255. Together with 
the low-ionization level of the system, this favors a picture where the cloud 
is closer to the IR source than to the UV source. This supports the idea that 
the extension of the IR source is larger than $\sim$200~pc. 
Alternatively, if the excitation is due to collisions, the electronic
density must be larger than 100~cm$^{-3}$. The dimensions of the
cloud along the line-of-sight is of the order of 0.006~pc and 
$\sim$50/$\sqrt k$ times larger in the perpendicular direction in order for the 
covering factor to be $\sim$97\% ($k$ is the amplification factor and
could be as large as 100).
Excited fine-structure lines are
not detected in the \zabs~=~3.9135 system. This suggests that the cloud is
farther away from the QSO compared to the \zabs~=~3.8931 cloud.
Using photo-ionization models with and without attenuation by 
associated systems acting as a screen between the QSO and the gas, it is 
shown that 
abundances are $Z\sim Z_\odot$ and 0.01~$Z_\odot$ at 
\zabs~=~3.8931 and  3.9135 respectively.
All this suggests that whereas the \zabs =  3.8931 system is probably
located within 200~pc from the QSO and ejected at a
velocity larger than 1000~km~s$^{-1}$, the \zabs = 3.9135 system must
be farther away and part of the host-galaxy. 

Aluminum could 
be over-abundant with respect to silicon and carbon at least by a factor of 
two and five at \zabs~=~3.8931 and  3.9135 respectively. Such enrichment has 
already been observed by Ganguly et al. (1999) in a system with a velocity 
separation of 10000~\kms~ with respect to Q1222+228.
It is well established that in Galactic halo stars and field dwarfs the 
odd-Z elements (Na, Al, Cl, P, Mn and Co) have metallicities relative to iron
smaller than solar by about a factor of 3 for iron metallicities in the range 
$-3.0~\le{\rm [Fe/H]~\le-1.0}$ (Timmes et al. 1995). Conversely, 
$\alpha-$chain elements (O, Ne, Mg, Si, S, Ar, Ca and Ti) generally have
metallicities relative to iron larger than solar by about a factor of three 
for [Fe/H]$\le-1.0$. The aluminum and silicon metallicities measured in 
damped Lyman-$\alpha$ systems follow these trends 
(Lu et al. 1997). In rapid star-formation models, massive star formation is 
favored and leads to high enrichment of aluminum and silicon with respect to 
carbon and nitrogen (Ferland et al. 1996). But [Si/H] is always larger than
[Al/H]. Large [Al/Fe] ratios are observed in bright giant stars in mildly 
metal-poor globular clusters (Ivans 1999). Various correlations between 
different element abundances suggest that the abundance pattern is most 
likely due to very deep mixing and proton-capture nucleosynthesis (see Langer 
et al. 1997 for a review). However, although [Al/Si] is found larger 
than solar, the excess is quite small ($<$~0.2~dex) and indeed much smaller 
than the factor 20 we observe in the $z_{\rm abs}$~=~3.9135 system.

Enrichment of aluminum with respect to silicon is found in the ejecta of 
classical novae (Andrea, Drechsel \& Starrfield 1994; Gehrz et al. 1998). 
Therefore, the over enrichment of odd nucleus could be explained 
if the absorption is produced by individual novae shells (Shields 1996). 
However, novae shells invariably produce very high metal enrichment and high 
values of [N/C] compared to solar (Petitjean et al. 1990; Andrea et al. 1994), 
which conflict with the present observations.

\vskip -0.5cm
\subsection{High-ionization systems}

The high-ionization associated systems at \zabs~=~3.90--3.917 (redshifts 
very close or even larger than the assumed intrinsic redshift, 
$z_{\rm em}$~=~3.91, see Fig.~8) have been shown to 
have metallicities larger than solar for a Mathews \& Ferland ionization
spectrum. At this redshift, the maximum 
life-time allowed by the $\Omega~=~1$ cosmology is 0.8 Gyr, suggesting that 
rapid star formation occurs. We note that the relatively low 
N~{\sc v} to \civ~ and \ovi~ column density ratios could be at odd with the 
standard rapid star-formation models discussed in the literature which 
predict larger values (Hamann \& Ferland 1993; Matteucci \& Padovani 1993).
However it is known that BALQSOs are X-ray quiet with optical
to X-ray spectral index $\alpha_{\rm OX}\ge 1.9$ (Green \& Mathur,
1996) when the Mathews \& Ferland spectrum corresponds to 
$\alpha_{\rm OX}~=~1.40$. If the cloud is ionized by a spectrum with
$\alpha_{\rm OX}\ge 1.9$, observations are consistent with metallicities
of the order of solar or slightly smaller.

The broad \civ~ absorption profile has a complex structure. It
shows mini-BAL absorptions (of width $\le 1000$~\kms) and narrow
components superposed on a continuous absorption of smaller optical
depth. There is a tendency for mini-BALs to have different covering
factors for different species. Again, this could be due to ionization
structure of the BLR or due to ionization gradients in the flow. It is
shown that the covering factors of a few of these absorbing clouds,
which are well detached from the corresponding broad emission line
profiles, are most probably due to the cloud not covering the three images of 
the lensed quasar. If true, this would put strong constraints on the
dimensions of the clouds. HST spectroscopic observations of the three
images are needed to confirm this findings. The continuous absorption is much 
stronger in \ovi~ indicating that the diffuse component is highly ionized 
compared to the other components in the flow. This suggests that the absorption
structures are due to density inhomogeneities within the flow.
 
The H~{\sc i} column density in the "mini-BAL" is small suggesting large 
metallicities as is commonly found in associated narrow as well as broad
absorption systems. However, there is no absorption from \aliii~
and P~{\sc v} contrary to what is seen in some of the BAL systems 
(Junkkarinen et al. 1995). Indeed, Shields (1997) derived 
overabundance of aluminum compared to silicon, [Al/Si] = 6[Al/Si]$_\odot$, 
in the BAL outflows of Q1101+091, Q1231+1325 and Q1331-0108, using
photo-ionization models with a Mathews \& Ferland (1987)  
spectrum and ionization parameters in the range $-$1.5~$<$~log~$U$~$<$~1.00. 
He also noted that there are indications for [Al/Si] being several 
times larger than solar in the BLR of some QSOs. 
The \aliii~ to \siiv~ optical depth ratio is less than 0.10 in the mini-BALs 
seen in the spectrum of APM~08279+5255. 
As the above ratio depends only weakly on the 
ionization parameter for log~$U~\ge~-0.50$ (Hamann 1997), we can use a simple 
scaling of the analysis of Shields (1997) to derive that 
[Al/Si]$\le$2 [Al/Si]$_\odot$.

Finally we tentatively identify {\sl narrow} components within the BAL-flow
which have velocity separations very close (within 5~\kms) to 
the \ovi, \nv~ and \siiv~ doublet splittings. This strongly suggests 
the existence of "line-locking" and points towards radiative acceleration as an
important process in driving the outflow.

\vskip -0.5cm

\acknowledgements{We would like to thank the team headed by Sara L. Ellison 
to have made this beautiful data available for general public use. 
We gratefully acknowledge support from the Indo-French Centre for the
Promotion of Advanced Research (Centre Franco-Indien pour la Promotion
de la Recherche Avanc\'ee) under contract No. 1710-1.}

\vskip -0.5cm

\end{document}